\documentclass[onecolumn,authoryear]{els-mrw} 

\usepackage{amsmath,amssymb,amsfonts,amsthm,makeidx,graphicx}
\usepackage{txfonts}
\usepackage{helvet}
\usepackage{aas_macros}
\usepackage[T1]{fontenc}

\newcommand\eg{{\it e.g.} }

\begin{document}

\chapter{White Dwarf Systems: the Composition of Exoplanets }\label{chap1}

\author[1]{Amy Bonsor}%

\address[1]{{Institute of Astronomy}, \orgdiv{University of Cambridge}, \orgaddress{Madingley Road, Cambridge CB3 0HA }}

\articletag{Chapter Article tagline: update of previous edition,, reprint..}

\maketitle

\begin{glossary}[Glossary]

\term{Accretion} is the ingestion of material by a star or other astronomical body

\term{Build-up phase} is the initial phase of accretion, where the abundances in the accreting material match those in the atmosphere

\term{Chondritic} refers to the composition of chondritic meteorites in the Solar System, whose refractory composition matches approximately that of the Sun

\term{Declining phase} refers to accretion that has finished but where metals are still visible as they sink out of the atmosphere

\term{Electron degeneracy pressure} is the quantum mechanical force that supports the interior of a white dwarf against gravitational collapse

\term{Iron core} is the metal core formed at the centre of a planet or planetary body following large-scale melting and metal-silicate segregation 

\term{{\bf {\it \bf Gaia}}} is a space-satellite launched by the European Space Agency in 2013 that monitors the positions, distances and motions of over 1 billion astronomical objects

\term{Large-scale melting} refers to a phase in the evolution of a planet when the planet is largely molten, often referred to as a magma ocean.

\term{Lithophile elements} (silicate-loving) have a strong affinity for oxygen and stay with the silicate during the large-scale melting that leads to core formation


\term{Metals} in the astronomical context refer to any element heavier than helium 

\term{Planetary nebula phase:} refers to the gentle release of the outer envelope of the asymptotic giant star

\term{Photosphere} refers to the region around a luminous object that is transparent to photons of a given wavelength

\term{Pollution:} is the term used to describe the presence of metals in the atmosphere of a white dwarf that should otherwise be pure hydrogen or helium

\term{Pollutants:} the planetary material in the atmosphere of white dwarfs

\term{Siderophile elements} (iron-loving) are species that move into the metal during melting

\term{Sinking}: refers to downwards trajectory of heavy elements due to gravity in the atmosphere of the white dwarf

\term{Spectra}: the intensity of light emitted over a range of energies or wavelengths

\term{Spectroscopy}: the study of spectra of astronomical objects to determine information regarding the material that leads to features in the spectra

\term{Steady-state accretion}: the phase of accretion where there is a steady-state between accretion into the top of the atmosphere and diffusion out of the visible atmosphere. 

\term{White dwarf:} the faint compact remnant leftover at the end of the lifetime of a star like our Sun

\end{glossary}



\begin{abstract}[Abstract]
We live in an exoplanet revolution, with more than 5,000 exoplanets detected to date. Our ability to characterise individual exoplanets is constantly improving, with exquisite mass and radius measurements for an ever-growing sample of planets, complimented by atmospheric characterisation of lower and lower mass planets. This chapter outlines a complimentary set of observations that uniquely provide bulk elemental compositions for exoplanetary material. Absorption features from metals, including Mg, Fe, Si, O, Ca, Al, Ni and Ti in the white dwarf photosphere characterise the composition of accreted planetary material. These observations highlight the diversity in composition across exoplanetary systems including volatile content and probe key geological processes including the formation of iron cores. Thanks to the many white dwarfs identified by the space satellite {\it Gaia}, a revolution in the spectroscopic characterisation of white dwarfs is underway.

\end{abstract}

\section{Contents}

\begin{enumerate}
\item{Contents}
    \item{Why are white dwarfs good tracers of exoplanet composition?}
    \item{Accretion and sinking}
    \item{Spectroscopic Observations as a means to trace planetary composition}
    \item{How common is planetary material in the atmospheres of white dwarfs?}
\item {White Dwarfs accrete predominantly rocky planetary material}
\item{Are the bodies accreted by some white dwarfs depleted in (moderate) volatiles?}
\item{Evidence for the formation of iron cores from white dwarfs}

\item{Trace elements}
Lithium, Potassium, Berryllium, 

{\item Conclusions and future outlook}
    
\end{enumerate}

\begin{BoxTypeA}[key]{Key Points:}
\begin{itemize}
    \item Spectroscopic observations of white dwarfs reveal metals from accreted planetary material
    \item Most white dwarfs accrete rocky material, dominated by Mg, Si, Fe and O, just like Earth, whilst some accrete material from water-rich asteroids or comets
    \item White dwarf planetary systems provide a way to probe the geology of exoplanetary systems
\end{itemize}
\end{BoxTypeA}

  \section{Introduction: Why are white dwarfs good tracers of exoplanet composition?}

White dwarfs are the faint remnants of low to intermediate mass stars like our Sun. More than 97\% of stars in the galaxy will end their lives as white dwarfs.
When main-sequence stars run out fuel, their outer envelopes swell to become giant stars. Eventually the outer layers of these stars become unbound and float off gently in a planetary nebula phase leaving behind a faint compact remnant. The compact remnant collapses under its own gravity until it reaches a sufficiently small size that it is supported by the electron degeneracy pressure. This faint, compact star cools as a white dwarf for the next 5-10Gyr. 

White dwarfs usually have carbon-oxygen cores, surrounded by a thin layer of helium and potentially hydrogen. A typical white dwarf has a mass of around 0.6$M_\odot$, but is about the size of the Earth. The strong gravitational forces mean that anything heavier than helium or hydrogen sinks out of sight in their thin atmospheres. Spectroscopic observations of white dwarfs should, therefore, reveal only features associated with hydrogen or helium.  If planetary material from a surviving outer planetary system enters the white dwarf atmosphere, this will show up as metallic features in the spectrum of the white dwarf. Comparison with models for the white dwarf atmosphere \cite[e.g.][]{Koester2010, Dufour2012} yields the relative abundances of many elements including key rock-forming elements such as Mg, Fe, Si and O, in the accreted planetary material. These spectroscopic observations of white dwarfs that have accreted planetary material uniquely provide the bulk elemental abundances of planetary material, in a manner which complements exoplanet detection and characterisation. Whilst exoplanet detections based on mass-radii measurements alone are often degenerate and atmospheric compositions are difficult to reconcile with surface conditions, white dwarf spectroscopy provides key elemental abundances including Mg/Fe or O/Fe.

\section{How does planetary material arrive into the atmospheres of white dwarfs?}
Exoplanets are prevalent around main-sequence stars. Although observations are biased towards close-in planets, a population of planets and planetesimal belts on wider orbits has already been found and in the future observatories, such as {\it Gaia}, {\it Roman} and {\it PLATO} will only increase our knowledge of these populations. Whilst inner planets are likely engulfed by the expanding giant star, anything beyond $\sim $1 au likely survives to the white dwarf phase. Stellar mass loss leads to an expansion of orbits. Thus, any surviving white dwarf planetary system will orbit far from the star. Common envelope evolution is one of two main methods which have been postulated to explain close-in planets observed around white dwarfs \citep[\eg][]{Lagos_2020}.

The other method is based on dynamical interactions between multiple bodies. Stellar mass loss increases the dynamical influence of planetary bodies onto one another. Many planetary systems may go unstable \citep{DebesSigurdsson},
whilst planets may scatter asteroids or comets. The presence of a binary stellar companion, asymmetric mass loss or planetary engulfment may also lead to dynamical scattering \citep{Hamers2016, Petrovich2017}. Any planetary body that is scattered close to the white dwarf, within what is known as the tidal radius, approximately a solar radius of the white dwarf, is torn to pieces by the strong tidal forces present \citep{Veras_tidaldisruption1}. 
Non-gravitational forces and collisional grinding within the highly eccentric ring of dust lead to accretion. Although a few gaps remain in our understanding of the exact accretion process here, radiative forces, most notably Poynting-Robertson drag can lead to circularisation and accretion of small dust particles \citep{Veras_review}. 

\begin{BoxTypeA}[chap1:box1]{Tidal Radius or Roche limit}
   \indent The distance from a star (or other body) where the tidal forces across a second body exceed its self-gravity. Any body that is held together purely by its own self-gravity will disintegrate. For white dwarfs, the tidal radius defines the distance from the star at which rubble pile asteroids, held together purely by their self-gravity, tidally disrupt. The Roche limit can be related to the ratio of the star (in this case white dwarf), density $\rho_{\rm WD}$ and the asteroid density $\rho_{\rm ast}$, for a spin-less rigid body:  
    \begin{equation}
        R_{\rm roche } = R_{\rm WD}\left(K \frac{\rho_{\rm WD}}{\rho_{\rm ast}}\right)^{1/3}, 
    \end{equation}
        where $R_{\rm WD}$ is the white dwarf radius and $K$ is a factor that depends on the rotation state and internal properties of the body. 
        
\end{BoxTypeA}

The accretion of planetary material by white dwarfs is witnessed by the observations of infrared emission from hot dust close to the star, circumstellar emission from metallic gas orbiting close to the white dwarf and the transits of planetary material seen for a handful of polluted white dwarfs \citep{Veras_review, Farihi_review}.


\begin{figure}[t]
\centering
\includegraphics[width=.9\textwidth]{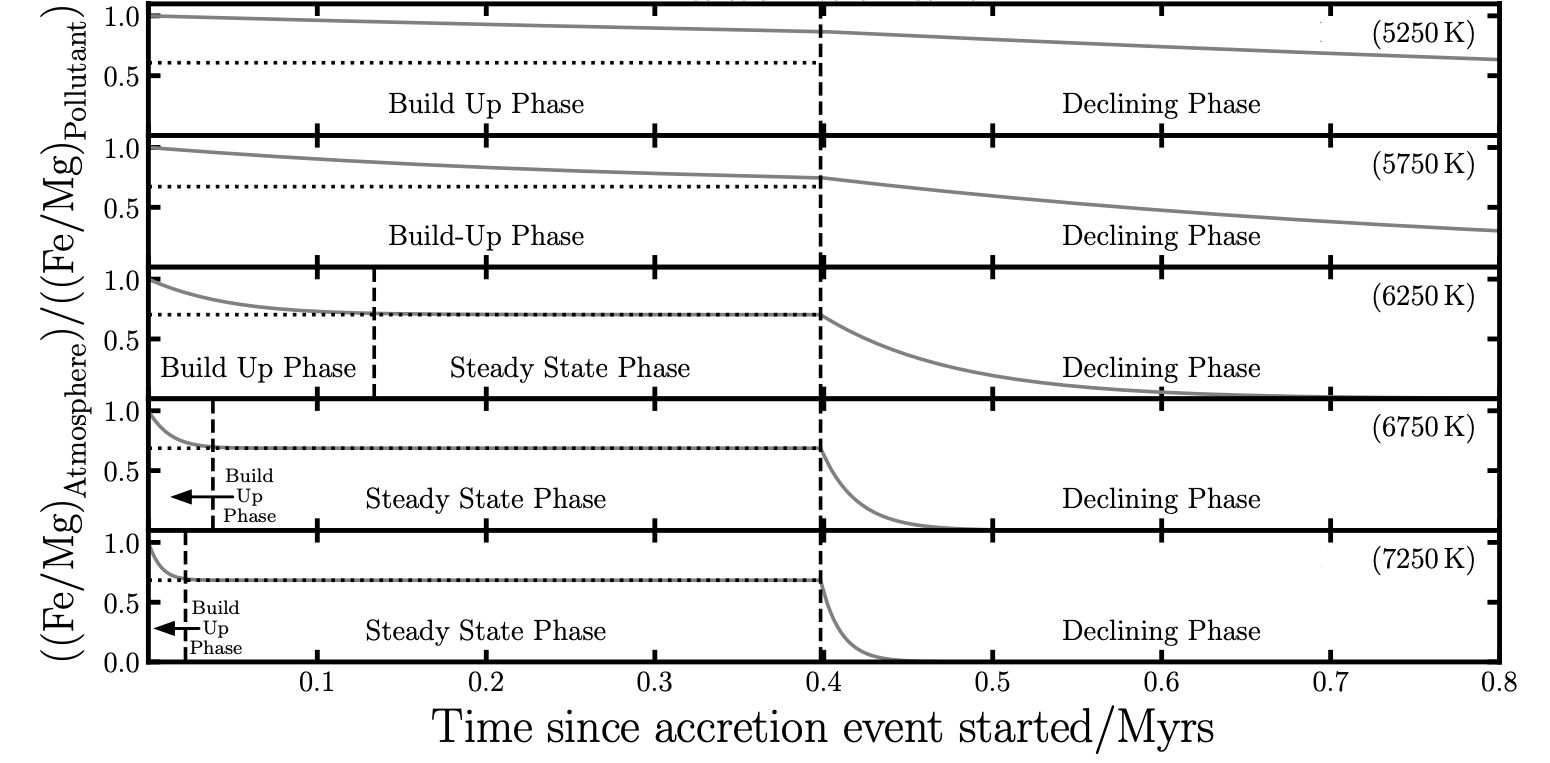}
\caption{The change in the relative abundances of elements observed in the atmosphere of a hydrogen-rich white dwarf (DA) relative to their relative abundances in the accreted planetary material (pollutant) depending on whether accretion has just started (build-up phase), accretion has reached a steady-state between accretion and diffusion, or accretion has finished (declining phase). The vertical dashed lines indicate the transition between accretion phases and the dotted lines represent the steady state solution given sufficiently long accretion. Each horizontal panel represents a different temperature white dwarf. Figure adapted from \citet{harrison_thesis}.}
\label{fig:sinking}
\end{figure}
  
    \section{Accretion and sinking}
The composition of the white dwarf atmosphere is modified by accretion of material and its atmospheric diffusion. Planetary material enters the top of the white dwarf atmosphere and sinks downwards due to the strong gravitational field of the white dwarf, eventually leaving the observable atmosphere. If accretion continues on long timescales, a steady-state between accretion into the top of the atmosphere and diffusion out of the base of the atmosphere is reached. The timescale on which each element sinks out of the visible atmosphere is defined as the sinking timescale. The elemental abundances (number ratios) observed in the atmosphere, $\left (\frac{El_1}{El_2}\right)_{atmos}$, may differ from those in the accreting material $\left (\frac{El_1}{El_2}\right)_{accreting}$, depending on the phase of accretion. The ratio between the abundances of two elements seen in the atmosphere is modified from those seen in the accretion in the following manner: 
\begin{equation}
    \left (\frac{El_1}{El_2}\right)_{atmos} = \left (\frac{El_1}{El_2}\right)_{accreting} \left (\frac{t_{sink, 1}}{t_{\rm sink, 2}}\right) \left ( \frac{ 1-e^{-t/t_{\rm sink, 1}}}{1-e^{-t/t_{sink,2}}}   \right). 
\end{equation}

When accretion initially starts in the {\bf build-up phase}, the abundances in the atmosphere match those of the accreting material. Once {\bf steady-state} is reached for both elements, the abundances in the atmosphere reach a steady-state in which they are modified by the ratio of the sinking timescales of the two elements in question. Once accretion finishes, the abundances {\bf decline}, such that the elemental ratios are now given by :
\begin{equation}
    \left (\frac{El_1}{El_2}\right)_{atmos} = \left (\frac{El_1}{El_2}\right)_{accreting} \left (\frac{t_{sink, 1}}{t_{\rm sink, 2}}\right)    
    \left ( \frac{ 1-e^{-t_{\rm event}/t_{\rm sink, 1}}}{1-e^{-t_{\rm event}/t_{sink,2}}}   \right)
    \, exp\left({-(t- t_{\rm event})}\left(\frac{t_{\rm sink, 2}- t_{\rm sink, 1}}{t_{\rm sink, 2}{t_{\rm sink, 1}}}\right )\right), 
\end{equation}
where $t_{\rm event}$ is the lifetime of the accretion event. The manifestation of this process can be seen in Fig.~\ref{fig:sinking}. 

\subsection{Radiative levitation}
In the hottest white dwarfs, the photon absorption cross-sections of particular ions are so high, that these species no longer gravitationally settle out of the atmosphere and rather are radiatively levitated \citep{Chayer1995}. Heavy elements are over-abundant in the atmospheres of white dwarfs hotter than 20,000K due to these effects. Radiative levitation must be considered when interpreting whether metals observed in the atmospheres of hot white dwarfs have a recent external origin.

    \section{Spectroscopic Observations as a means to trace planetary composition}

\begin{figure}[t]
\centering
\includegraphics[width=.8\textwidth]{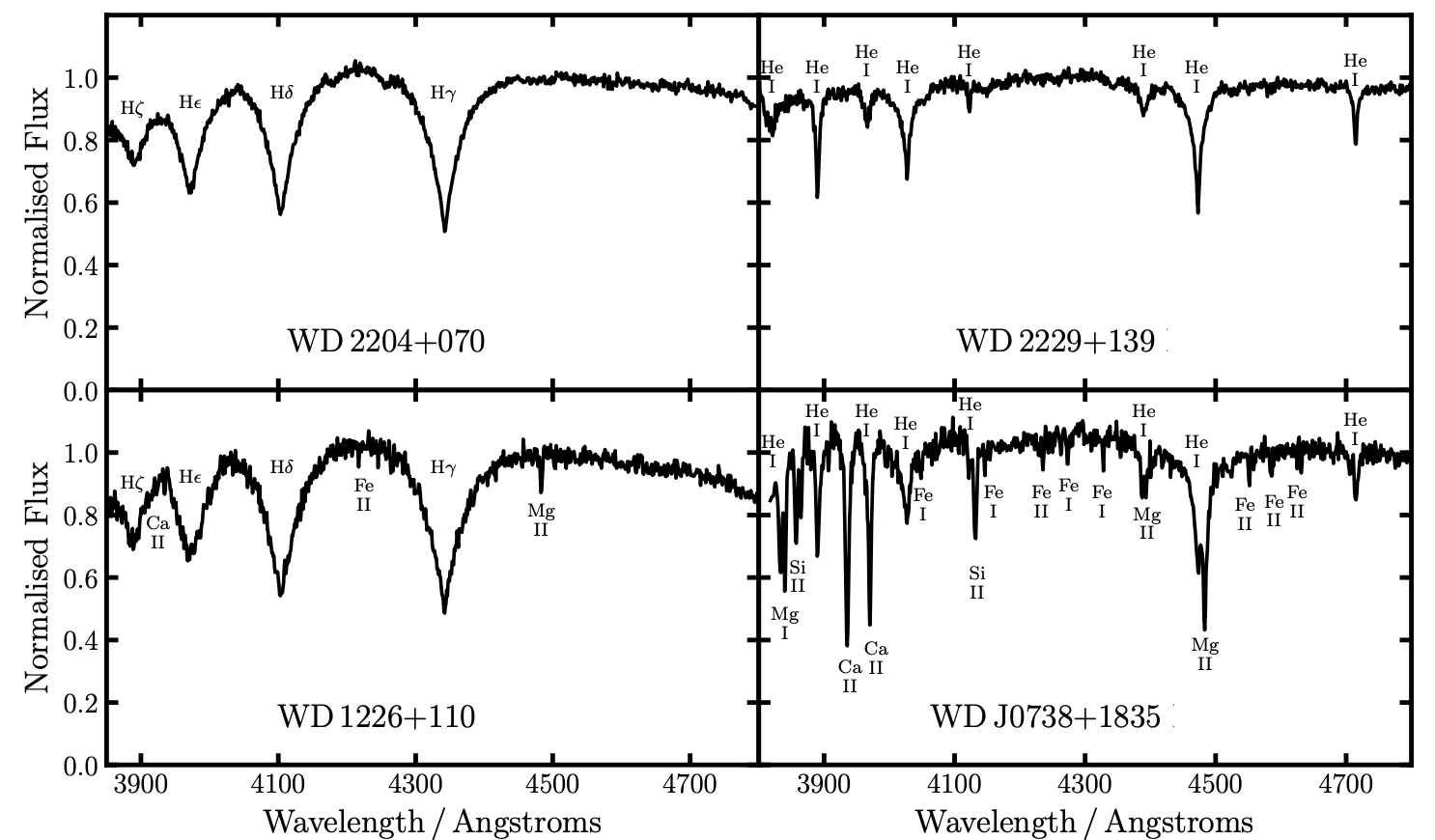}
\caption{The spectra of four white dwarfs in the region $3800\AA$ to $4800\AA$, highlighting key differences between hydrogen-dominated (right) and helium- dominated (left) atmospheres, where the atmosphere contains metals (bottom) and does not contain metals (top) \cite{harrison_thesis}. }
\label{fig:spectra}
\end{figure}

The radiation from a star, including a white dwarf, is to first approximation, the same as that from a black-body of a given temperature. White dwarfs are in general, UV bright, hot stars. As radiation escapes from the central star, atomic gases in the atmosphere absorb radiation at particular wavelengths. These absorption features leave a particular fingerprint in the white dwarf spectra. For most white dwarfs, this absorption is dominated by hydrogen, with strong Balmer features, leading to the typical spectrum of a DA (hydrogen-dominated) white dwarf, as shown in the upper-left panel of Fig.~\ref{fig:spectra}. For those white dwarfs with helium dominated atmospheres (DBs), the Helium absorption features dominate the spectrum (upper right panel of Fig.~\ref{fig:spectra}). 

If metals are present in the atmosphere of a white dwarf, for example due to the accretion of planetary material, absorption features due to these metals show up in the atmosphere, as shown in the bottom panels of Fig.~\ref{fig:spectra}. Most commonly, the strong features from Calcium (Ca H $\&$ K lines at $3934\AA$ and $3969\AA$ and Ca infrared triplet at $8498\AA$, $8542\AA$ and $8662\AA$). The exact combination of metal features depends on: 
\begin{itemize}
    \item the abundances of each element
    \item the strength of the atomic features
    \item the temperature of the white dwarf 
    \item the depth of its visible atmosphere.  
\end{itemize}

\subsection{Elemental abundances from optical and UV spectra}
Whilst optical spectra from the ground are very powerful in probing the atmospheric composition of white dwarfs, there are a range of elements that only have absorption features accessible in the UV, necessitating the need for space-based observatories. Elements such as C, S, P, N are predominantly found from their features in UV spectra. Elements such as O and Si occur in both UV and optical spectra. This poses an important open question for the field, as there is commonly a discrepancy found between abundances derived from optical and UV spectra for the same elements \citep[\eg][]{Xu2019}. \cite{Rogers2024} speculate that this discrepancy could be related to the depth of formation of the spectral lines. Caution must, thus, be used when considering the full abundances of the material and it is advisable to compare abundances derived from UV spectra only to other abundances determined from UV spectra. 
    
    \section{How common is planetary material in the atmospheres of white dwarfs?}

This is a crucial question, not only in ascertaining the viability of utilising white dwarf spectra to probe planetary composition, but in determining how commonly exoplanetary systems, and the mechanisms that lead to accretion, exist around white dwarfs.

\begin{figure}[t]
\centering
\includegraphics[angle=0, width=.8\textwidth]{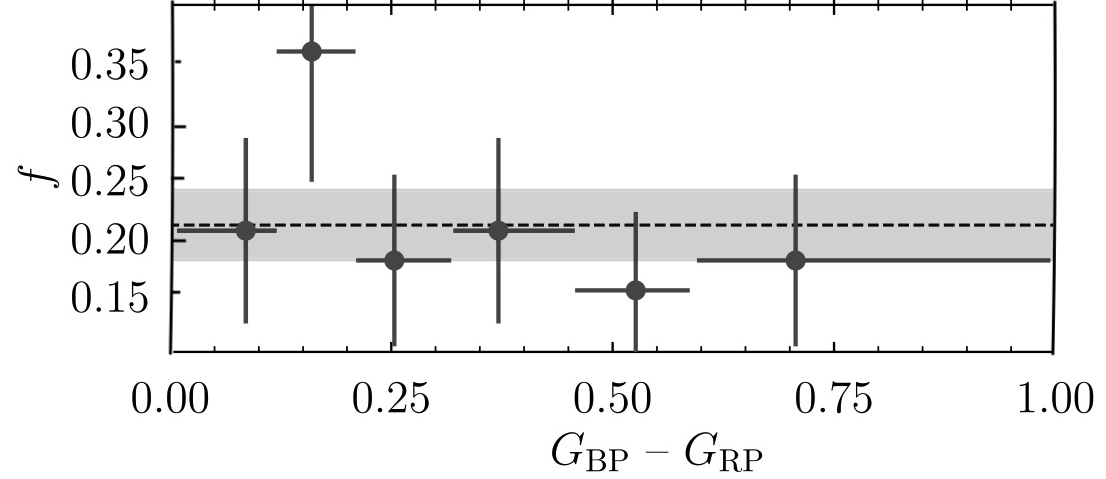}
\caption{ The  the fraction (21\%) of helium-dominated white dwarfs that are metal enriched as a function of {\it Gaia} colours, G$_{\rm BP}$-G$_{\rm RP}$ in DESI spectra \citep{Manser2024b}. }
\label{fig:manser}
\end{figure}

Spectroscopic observations tells us that planetary systems are common around white dwarfs. Exactly how common is harder to resolve, as there is a strong dependence of the ability to detect metals on the properties of the white dwarf, as well as the resolution and signal-to-noise ratio of the spectra. \cite{Zuckerman03, Zuckerman2010} found 25\% (30\%) of single DA (DB) white dwarfs contain Ca. For hotter stars  ($17 000$ K$ <T_{\rm eff} < 27, 000$ K), \cite{Koester2014} use the Hubble Space Telescope to find Si in 56\% of a sample of 85 DA white dwarfs, with 25\% requiring recent accretion. Based on the white dwarfs identified by {\it Gaia} \cite{Obrien2024} find Ca in 11\% of white dwarfs in the 40pc sample, whilst \cite{Manser2024b} find metal enrichment in  $21\pm 3$ \% of a sample of 234 white dwarfs observed with the Dark Energy Spectroscopic Instrument (DESI), as shown in Fig.~\ref{fig:manser}.

\section {White Dwarfs accrete predominantly rocky planetary material}

\begin{figure}[t]
\centering
\includegraphics[angle=270, width=.7\textwidth]{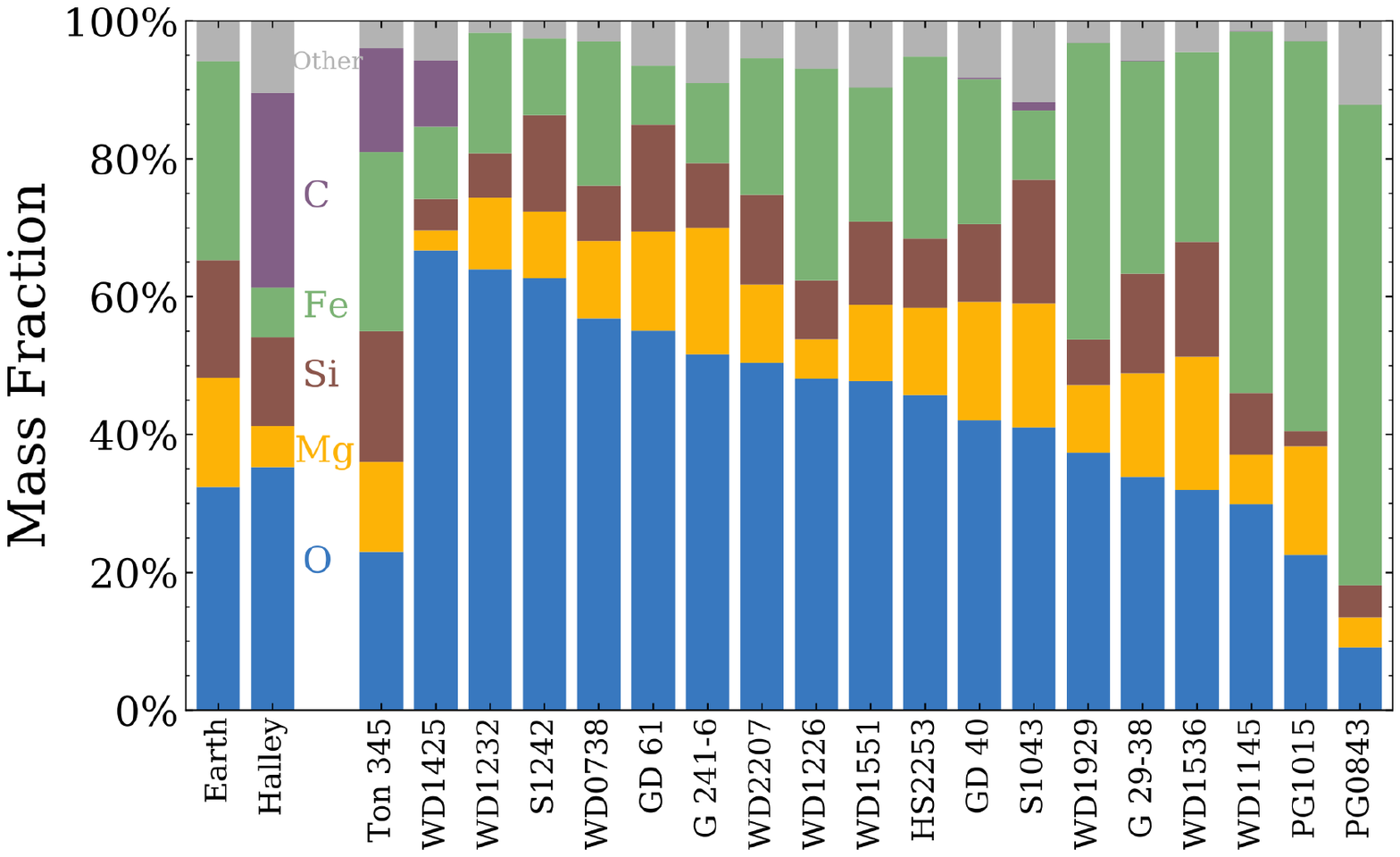}
\caption{ Mass fractions of different elements—O, Mg, Si, Fe, C, and other—accreted onto polluted white dwarfs. The compositions of bulk Earth and comet Halley are also shown for comparison at left. Most white dwarfs (with their name designations given along horizontal axis) appear to have accreted rocky planetesimals \citep{XuBonsor2021Elements}. }
\label{fig:abundances}
\end{figure}

\begin{figure}[t]
\centering
\includegraphics[angle=0, width=.8\textwidth]{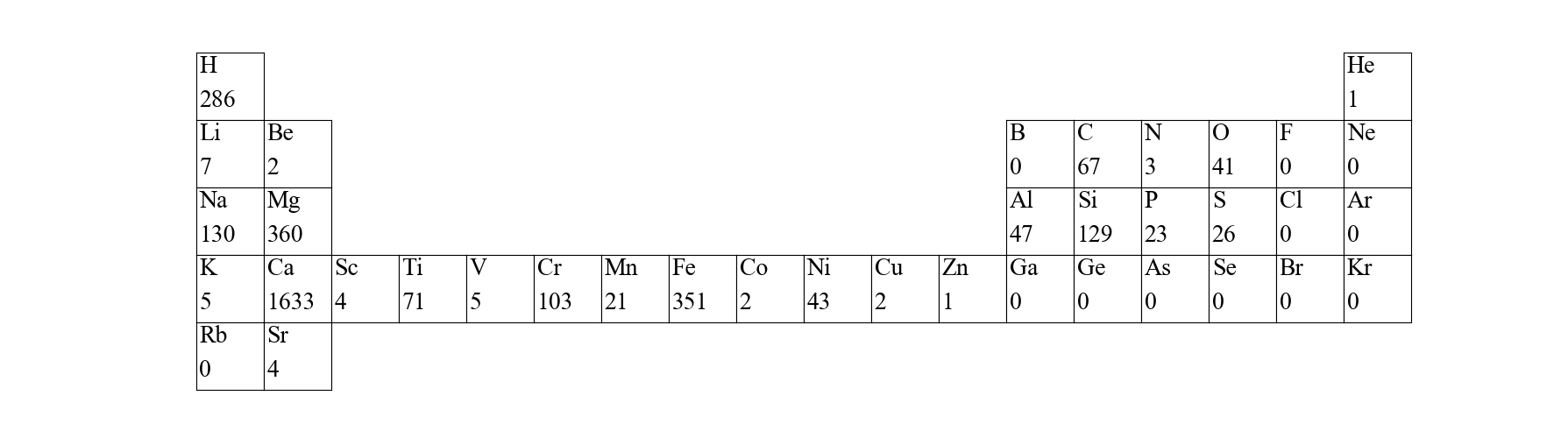}
\caption{A diverse range of elements have now been detected in the atmospheres of white dwarfs. This figure shows the number of objects detected with each element on the periodic table (Williams et al, in prep).  }
\label{fig:williams}
\end{figure}

Whilst Calcium is the element most commonly detected in white dwarf atmospheres, the relative abundances of all four major rock-forming elements- Magnesium, Iron, Oxygen and Silicon- can be found in the material accreted by a substantial subset of white dwarfs. Fig.~\ref{fig:abundances} shows the diversity in Mg, Fe, O and Si composition. Bulk Earth is shown for comparison and whilst most white dwarfs exhibit similar bulk compositions to Earth \citep{Trierweiler2023}, there is diversity in the exact ratios of the major-rock forming elements, in particular, both the oxygen and iron content. The compositions are, however, clearly distinct from stellar (or inter-stellar) compositions, with the material accreted by white dwarfs showing sub-solar volatile content, notably O, but also C, N, Na. A wide range of different elements have now been detected in the atmospheres of white dwarfs. Fig.~\ref{fig:williams} shows the number of  white dwarfs with different elements detected across the periodic table. As of 2024, over 1,600 white dwarfs are known to have metals in their atmospheres.

\section{Do white dwarfs accrete comets or asteroids?}
In the Solar System, there is a clear dichotomy in the volatile content pertaining to the two major groups of small bodies. In the simplest possible terms comets are icy and asteroids rocky. In reality there is a spectrum of volatile content across the Solar System, with in particular carbonaceous meteorites, potentially from the outer-asteroid belt, containing significant volatile budgets. 

\begin{figure}
    \centering
    \includegraphics[width=.5\textwidth]{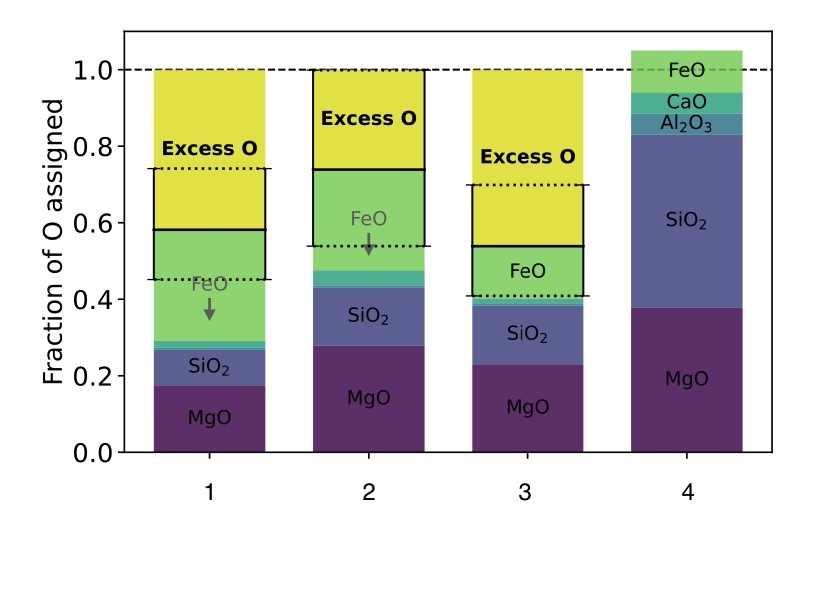}
    \caption{The total oxygen observed in the atmosphere of four white dwarfs, labelled 1 to 4, is likely divided between metal oxides (MgO, SiO$_2$, Al$_2$O$_3$, CaO, and FeO) and water-ice, labelled as excess oxygen (yellow). The first three white dwarfs shown have sufficient oxygen that it was accreted not only as metal oxides, whilst the full oxygen budget of the fourth white dwarf can be explained by the metal oxides alone. The error range results from measurement errors and an uncertainty as to whether Fe exists as Fe or FeO \citep{Rogers2024b}.  }
    \label{fig:oxygen_excess}
\end{figure}

In a similar manner, the bodies accreted by white dwarfs sample a range of volatile contents. Purely based on their oxygen abundances, relative to the other metals present, it seems that a handful of white dwarfs accreted planetary material with a significant ice budget. Whilst most oxygen is locked up in metal oxides, water is the most likely carrier for the remaining or `excess' oxygen. Fig.~\ref{fig:oxygen_excess} shows that for some white dwarfs, more oxygen is accreted than can be accounted for in rocks. 

This presents a puzzle, as the planetary bodies that orbit white dwarfs have survived intense periods of irradiation during the star's evolution on the giant branches \citep{jura2010}. \cite{Malamud2017} suggest that full water retention only occurs for bodies orbiting at hundreds of au around massive stars, but partial water retention can occur much closer. The water-rich bodies accreting onto some white dwarfs must originate from the very outer regions of planetary systems, potentially even exo-Oort clouds. 

Whilst it is relatively hard for any material from a disrupted planetary body to leave the white dwarf planetary system, as the radiation from the white dwarf is insufficient to expel material, \cite{Brouwers2023_async_volatiles} suggest that volatiles may be accreted asynchronously to refractory material, with the accretion of a single planetary body undergoing periods of volatile-rich and refractory-rich accretion.

Water contains not only oxygen but hydrogen. For most white dwarfs this hydrogen is indistinguishable from the hydrogen dominating the atmospheric composition. However, some helium-rich white dwarfs contain trace hydrogen. It has been suggested that a natural explanation for this hydrogen is the accretion of icy or water-rich comets \citep{GentileFusillo2017}. In which case, the total hydrogen traces the history of accretion of icy bodies over the lifetime of that white dwarf.

\begin{BoxTypeA}[chap1:box1]{Oxygen Excess}
\indent Excess oxygen refers to the oxygen in the planetary material that is not locked up in metal oxides, including MgO, CaO, SiO$_2$, Al$_2$O$_3$. This oxygen is most likely carried as water-ice, H$_2$O, which could be in the form of pure water-ice or hydrated minerals. In order to calculate whether any water is present in the material accreted by a white dwarf, the total number of oxygen atoms accreted is allocated to each metal:

\begin{equation}
    N_{O, excess} = N_O- N_{Mg}- 2N_{Si} -N_{Ca} - 2/3 N_{Al}\, \,  - X\, N_{Fe}. 
\end{equation}
It is not clear whether iron is present as iron metal, Fe ($X=0$), iron oxide FeO ($X=1$) or Fe$_2$O$_3$ ($X=3/2$), thus, there always remains some uncertainty in the oxygen excess. 
\end{BoxTypeA}

\section{Evidence for the formation of iron cores from white dwarfs}
Core formation is a key geological process, providing evidence for a period of large-scale melting during the history of a planetary body. Iron cores have a distinct compositional signature compared to silicate mantles. Some material accreted by white dwarfs seems to be enhanced in siderophile (iron-loving) elements compared to lithophile (silicate- loving) elements suggesting that the material currently being accreted is dominated by the core of the original planetary body \citep{Melis2011, Hollands2018}. For others, the opposite is true and the material appears dominated by the mantle \citep{Swan2019, Buchan2022}, or even crustal material \citep{Zuckerman2011}. 
The white dwarf may currently be witnessing a portion of the entire body being accreted, with some models suggesting that the accretion of a single planetary body occurs via a phase of core-rich followed by mantle-rich composition \citep{Brouwers2023_async}. Alternatively, the white dwarf may be accreting an entire planetary body, which is itself a collision fragment of a larger planetary body, collisionally eroded during the billions of years of evolution in the outer planetary belt, on the way to the white dwarf phase. White dwarfs that accrete material from both siderophile and lithophile elements are uniquely placed to assess how prevalent the formation of iron cores in exoplanetary building blocks might be \citep{bonsor2020} and following on from this, the timing of planetesimal formation \citep{Bonsor_al26}.

\begin{BoxTypeA}[chap1:box2]{Iron Core Formation}
\indent Large-scale melting can be fueled by the gravitational potential energy released during the accretion of a planetary body (for diameters $>$1,000km or due to heating from the decay of radio-active nuclides. In the Solar System, $^{26}$Al (half-life 0.7Myr) led to melting in bodies larger than 10km that formed earlier than $\sim$1Myr. Within a molten body, iron droplets form, which sink under gravity to form an iron core, leaving behind a silicate mantle. The exact composition of these iron droplets depends on the conditions in the molten body. Siderophile species (iron-loving) enter the iron droplets and form the core, whilst lithophile species remain in the silicate. 
   
\end{BoxTypeA}

\section{Trace elements}

A wide variety of elements have now been detected in the atmospheres of white dwarfs. The abundances of many of these elements provide key insights regarding the geophysical evolution of the planetary bodies accreted by these white dwarfs. A few examples include:

\indent \indent {\bf Nickel, Chromium}: this pair of siderophile elements are good tracers of iron core formation and the conditions under which the core formed. \cite{Buchan2022} postulate that Ni/Fe and Cr/Fe have the potential to constrain the size of the original planetary body when the iron core formed, if a white dwarf can be found that accreted material that is sufficiently core (mantle)-rich. 

{\bf Alkali elements, lithium, potassium, sodium}: all have deep absorption features in cool, metal enriched white dwarfs and many new objects have recently been found \citep{Kaiser2021, Hollands2021}. K, Li and Na are all important tracers of crust formation. Li is an important tracer of chemical enrichment across the galaxy.  \cite{Kaiser2021} postulate that white dwarfs enriched in Li have an advantage over stars, where several out-standing puzzles regarding Li abundances remain.

{\bf Moderate volatiles, Manganese, Sodium:}
Na is commonly detected in cool white dwarfs \citep[\eg][]{Hollands2017}, whilst Mn has strong lines in the UV \citep[\eg][]{Gaensicke2012}. Together these two elements trace volatile loss in planetary bodies. Many planetary bodies experience a phase of impact induced melting. This occurs under oxidising conditions, which render Na more volatile relative to Mn. This has a distinct signature to volatile loss, as the first solids condense out of the nebula gas. \cite{Harrison_MnNa} use Mn and Na abundances in four white dwarfs to show that whilst most planetary material accreted by white dwarfs has abundances determined early in planet formation (incomplete condensation), the body accreted by GD 362 likely underwent a phase of violent impacts.  


{\bf Berryllium:} has now been found in the atmospheres of two white dwarfs GALEX J2339 and GD 378 \citep{Klein2021}, where the abundances were two orders of magnitude higher than chondritic. Such high abundances have been attributed to spallation of either C, O or N, potentially on an exo-moon \citep{Doyle2021}. 

{\bf Carbon:} has strong lines in the UV and has been found in tens of white dwarf atmospheres. Carbon and oxygen together provide key tests for planet formation models. No white dwarf seems to accrete material that is purely carbon-rich and white dwarfs accrete material with C/O that is broadly sub-solar, similar to Solar System rocky bodies \citep{Wilson2016, Rogers2024}.

{\bf Nitrogen:} has been detected in the atmosphere of a single white dwarf, WD 1425+540 \citep{Xu2017} to date. Nitrogen is very volatile and depleted (relative to solar) even in Solar System comets. This white dwarf likely accreted material from a Kuiper-belt analogue, also rich in C and O.

\section {Conclusions and Future Outlook }

The {\it Gaia} space satellite has prompted a revolution in many fields of astronomy, with the study of white dwarfs being no exception \citep{Tremblay2024}. Precise {\it Gaia} parallaxes have enabled hundreds of thousands of white dwarfs to be identified \citep{Gentile-Fusillo2021}. Many of these white dwarfs are readily accessible to spectroscopic follow-up from the ground. These spectroscopic observations are already starting to yield many exciting new results across the fields of white dwarf evolution, galactic evolution and exoplanetary science.  \cite{Manser2024, Manser2024b} presented more than a thousand spectra from {\it DESI}. 4MOST, SDSS-VII and WEAVE all promise to provide exciting results in the next few years.

\begin{ack}[Acknowledgments]
\indent AB acknowledges support
of a Royal Society University Research Fellowship, URF$\backslash $R1$\backslash $211421. AB is grateful to Yuqi Li, Laura Rogers and Wei Qiang for helpful comments that improved the quality of the manuscript. 

\end{ack}

\seealso{article title article title}

\bibliographystyle{Harvard}
\bibliography{ref}

\end{document}